\documentclass[twocolumn,showpacs,preprintnumbers,amsmath,amssymb]{revtex4}

\usepackage{graphicx}
\usepackage{dcolumn}
\usepackage{bm}
\usepackage{amsthm}
\begin{document}


\title{Capability of local operations and classical communication for distinguishing bipartite unitary operations}

\author{Lvzhou Li$^{a, c}$}\email{lilvzh@mail.sysu.edu.cn
(L. Li).} \author{Shenggen Zheng$^a$} \author{Haozhen Situ$^b$} \author{Daowen Qiu$^a$}

 \affiliation{%
 $^a$ Institute of Computer Science Theory, School of Data and Computer Science, Sun Yat-sen University, Guangzhou 510006,
 China
}%

 \affiliation{%
 $^b$ College of Mathematics and Informatics,
  South China Agricultural University,
Guangzhou 510642, China
}%

\affiliation{%
 $^c$ The Key Laboratory of Machine Intelligence and Advanced Computing (Sun Yat-sen University)
Ministry of Education, China
}%

\date{\today}

\begin{abstract}
The problem behind this paper is, if the number of queries to unitary operations is fixed, say $k$, then when do  local operations and classical communication (LOCC) suffice for  optimally distinguishing bipartite unitary operations?  We consider the above problem for two-qubit unitary operations  in the case of $k=1$, showing that for two two-qubit entangling unitary operations without local parties,  LOCC  achieves the same distinguishability  as the global operations. Specifically, we obtain:  (i) if such two unitary operations are perfectly distinguishable by global operations, then they are perfectly distinguishable by LOCC too, and (ii) if they are not perfectly distinguishable by global operations, then  LOCC can achieve the same optimal discrimination probability  as  the global operations.
\end{abstract}

  \maketitle

\section{Introduction}
\label{sec:intro}

Distinguishability, as a fundamental  concept, lies at the heart of  quantum information theory, with a wide range of applications in quantum information and computation. While distinguishability of quantum states has been intensively and extensively studied,  it has also been extended to quantum evolution  in various forms such as distinguishability of unitary operations \cite{Aci2001,MAP2001,DFY2007,Zhou2007,DFY2008,LQ2008, Li2017,CAR2013, Bae2015,Gao2016}, measurements \cite{Ji2006}, Pauli channels \cite{DSK2005},   oracle operators \cite{CKTS2007}, and quantum operations \cite{Sac2005, LQ2008JPA, WY2006,Wat2008,DFY2009}. In this paper, we focus on  distinguishability of unitary operations.

Discrimination of  unitary operations is generally transformed to discrimination of quantum states by preparing a probe state and then discriminating the output states generated by different unitary operations.  Two unitary operations $U$ and $V$ are said to be
perfectly distinguishable (with  a single query), if there
exists a state $|\psi\rangle$ such that $U|\psi\rangle\perp
V|\psi\rangle$.   It has been  shown that  $U$ and
$V$ are perfectly distinguishable
if, and only if $\Theta(U^\dagger V)\geq\pi$, where $\Theta(W)$
denotes the length of the smallest arc containing all the
eigenvalues of $W$ on the unit circle \cite{Aci2001,MAP2001}. The situation changes dramatically when  multiple queries  are allowed, since any two different unitary operations are perfectly distinguishable in this case. Specifically,
it was shown that for any two different unitary
operations $U$ and $V$, there exist a finite number $N$ and a
suitable state $|\varphi\rangle$ such that $U^{\otimes
N}|\varphi\rangle\perp V^{\otimes N}|\varphi\rangle$  \cite{Aci2001,MAP2001}.
Such a discriminating scheme is intuitively called a {\it parallel scheme}. Note that in the parallel scheme, an $N$-partite
entangled state as an input is required  and plays a crucial role. Then, the
result was further refined in Ref. \cite{DFY2007} by showing that the
entangled input state is not necessary for perfect discrimination of unitary operations. Specifically, Ref. \cite{DFY2007} showed
that for any two different unitary operations $U$ and $V$, there
exist an input state $|\varphi\rangle$ and auxiliary unitary operations
$w_1,\dots,w_N$ such that $Uw_NU\dots w_1U|\varphi\rangle\perp
Vw_NV\dots w_1V|\varphi\rangle$.  Such
a discriminating scheme is generally called a {\it sequential scheme}.

Note that in these researches mentioned above, it was assumed by default that the  unitary operations to be discriminated are under the
complete control of a single party who can  perform any physically
allowed operations to achieve an optimal discrimination. Actually, a more complicated case is that the
 unitary operations to be discriminated   are shared by
several spatially separated parties. Then, in this case a reasonable constraint
on the discrimination is that each party can only make local
operations and classical communication (LOCC). Despite this constraint, it has been shown that  any two bipartite unitary
operations can be perfectly discriminated by LOCC, when   multiple  queries to the unitary operations are allowed \cite{Zhou2007,DFY2008,LQ2008, Li2017}. More specifically, Refs. \cite{Zhou2007,DFY2008} independently proved this result with tools from  universality of quantum gates \cite{Bry} and  analysis of
numerical range \cite{Horn}, respectively.  However,   Refs. \cite{Zhou2007,DFY2008} generally required a complicated network  combining the sequential and the parallel schemes to achieve a perfect discrimination, where one of the two parties who share the bipartite unitary operations must prepare  a multipartite entangled state.

A further result was  obtained in Ref. \cite{LQ2008} which asserts that any two bipartite unitary operations acting on  $d\otimes d$ (i.e., a two-qudit system) with multiple queries allowed, in principle, can be perfectly discriminated  by LOCC  with  merely a sequential scheme. Note that a sequential scheme usually represents the most economic strategy for discrimination, since it does not require any entanglement and saves the  spatial
resources.  Nevertheless, the result in Ref. \cite{LQ2008} has two limitations:  (i) the  unitary operations  to be discriminated were limited to act on $d\otimes d$, and (ii) the inverses of the unitary operations  were assumed to be  accessible, although this assumption may be unrealizable in experiment. Therefore, the first author improved the result  in Ref. \cite{Li2017} by showing that any two bipartite unitary operations acting on $d_A\otimes d_B$ can be locally discriminated with a sequential scheme, without using the inverses of the unitary operations.

After these work, we have a relatively comprehensive understanding on local discrimination of  unitary operations. The above results imply that  LOCC and global operations can achieve the same distinguishability for unitary operations---perfect discrimination for both two cases,  when the  unitary operations can be queried multiple times. But, note that for achieving a perfect discrimination, the two situations may  require different numbers of queries  to the unitary operations. Let $N$ be the optimal number of queries to $U$ and $V$  for a perfect discrimination between them in the case of global operations (similarly, $N'$ is the one for the case of LOCC). Then, it is obvious that $N^{'}\geq N$. However,  what is the  condition for $N^{'}=N$  seems  unknown until now. Thus, this inspires us to consider such a question: if the number of queries to unitary operations is fixed, say $k$, then when do LOCC suffice for optimally distinguishing bipartite unitary operations?

In this paper, we consider the above problem for distinguishing  two-qubit unitary operations  in the case of  $k=1$ (i.e., the unitary operations can be queried only  once).  We  show that if  two two-qubit entangling unitary operations without local parties can be queried only once, then LOCC  achieve the same distinguishability as the global operations. More specifically, we obtain:  (i) if the two unitary operations are perfectly distinguishable by global operations, then they are perfectly distinguishable by LOCC too, and (ii) if they are not perfectly distinguishable by global operations, then  LOCC can achieve the same optimal discrimination probability as  the global operations. We hope these discussions about this elementary case would shed some light on the more generalized cases.

The main idea of our method is described as follows. First, the error probability of discriminating between $U_1$ and $U_2$ is given by
$$P_E(U_1, U_2)=\frac{1}{2}\left(1-\sqrt{1-4p_1p_2F(U_1, U_2)^2}\right),$$ where $F(U_1, U_2)=\min_{|\psi\rangle}|\langle\psi| U_1^\dagger U_2|\psi\rangle|$.
If  $U_1$ and $U_2 $ are acting on a two-qubit system $AB$, and we want to discriminate them by LOCC, then the probe state $|\psi\rangle$ should be a product state, that is, $|\psi\rangle=|\psi_A\rangle\otimes |\psi_B\rangle$. Note that for discriminating two multipartite states, it has been shown that LOCC can achieve the same distinguishability that the global operations would have \cite{Wal2000,Vir2001}. Therefore, if we can find a product state  $|\psi\rangle$ such that $F(U_1, U_2)=|\langle\psi| U_1^\dagger U_2|\psi\rangle|$, then it can be asserted that   LOCC as powerful as the global operations in discriminating $U_1$ and $U_2$. We will prove this point by using some simple  geometric knowledge.
Note that Ref. \cite{Bae2015} obtained a similar result. However, if one carefully checks the result there, then it could be found that the result in this paper seems more generalized and different ideas are used to derive the results.

The rest of this paper is organized as follows. Section \ref{sec:1} recall the canonical decomposition of two-qubit unitary operations. The main result   is presented in Section \ref{sec:2}. A conclusion is made in Section \ref{Con}.



\section{Decomposition of two-qubit unitary operations\label{sec:1}}
Concerning decomposition of two-qubit unitary operations and  related notations, one can refer to Ref. \cite{Kraus2000} and references therein for details, and here we only recall some  necessary results. Any  unitary operation $U$ acting on two qubits $A$ and $B$  has the following canonical decomposition:
\begin{align}
U=(U_A\otimes U_B)U_d(V_A\otimes V_B),\label{U}
\end{align}
where  $U_A$, $U_B$, $V_A$ and $V_B$ are single-qubit unitary
operations and $U_d$ has the following form
\begin{align}
U_d=e^{-i(\alpha_x\sigma_x\otimes\sigma_x+\alpha_y\sigma_y\otimes\sigma_y+\alpha_z\sigma_z\otimes\sigma_z)}. \label{Decomposition1}
\end{align}
Here, $\sigma_x$, $\sigma_y$, and $\sigma_z$ are Pauli operators,
and the  vector $d=(\alpha_x, \alpha_y, \alpha_z)$ has real
entries satisfying
\begin{align*}
0\leqslant\alpha_z\leqslant
\alpha_y\leqslant\alpha_x\leqslant\frac{\pi}{4}.
\end{align*}

If $\alpha_x=\alpha_y=\alpha_z=0$, then $U_d=I$.  Else if $\alpha_x=\alpha_y=\alpha_z=\frac{\pi}{4}$, then $U_d$ is  the  SWAP operation. Otherwise, it is {\it entangling}, that is,  it can create entanglement between two qubits initially in a product state. Thus, $U_d$ is called the {\it entangling part}  of $U$ in this paper.  Denote the  Bell states by $|\Phi^{\pm}\rangle=(|00\rangle\pm |11\rangle)/\sqrt{2}, |\Psi^{\pm}\rangle=(|01\rangle\pm |10\rangle)/\sqrt{2}$. The following states form the so-called magic basis:
\begin{align*}
&|\Phi_1\rangle=|\Phi^+\rangle, ~~ |\Phi_2\rangle=-i|\Phi^-\rangle,\\
&|\Phi_3\rangle=|\Psi^{-}\rangle, ~~ |\Phi_4\rangle=-i|\Psi^{+}\rangle.
\end{align*}
Then,  $U_d$ is diagonal in
the magic basis, and it can be  written as
  \begin{equation}
U_d=\sum_{j=1}^{4}e^{-i\lambda_j}|\Phi_j\rangle \langle\Phi_j|,
\label{Decomposition2}
\end{equation}
where
\begin{align*}
\lambda_1=\alpha_x-\alpha_y+\alpha_z, ~~ \lambda_2=-\alpha_x+\alpha_y+\alpha_z, \\
\lambda_3=-\alpha_x-\alpha_y-\alpha_z, ~~\lambda_4=\alpha_x+\alpha_y-\alpha_z.
\end{align*}
It is easily seen that $\lambda_4\geq \lambda_1\geq \lambda_2\geq \lambda_3$.

Any two-qubit state $|\psi\rangle$ can be represented as $|\psi\rangle=\sum_k u_k|\Phi_k\rangle$, where $\sum_k |u_k|^2=1$. We can use the so-called {\it concurrence} $C$ to measure the entanglement of a pure state of two qubits, which is defined by
$$C(|\psi\rangle)=|\langle\psi|\sigma_y\otimes \sigma_y|\psi^*\rangle|,$$
where $|\psi^*\rangle$ denotes the complex conjugate of $|\psi\rangle$ in the computational basis. Writing $|\psi\rangle$ in the magic basis, we get
 $$C(|\psi\rangle)=\left|\sum_k u^2_k\right|.$$

Thus, $C(|\psi\rangle)=0$, that is,  $|\psi\rangle$ is a product state,  iff $\sum_k u^2_k=0$.

\section{Discrimination of entangling two-qubit unitary operations\label{sec:2}}

Note that when considering discrimination of quantum states or quantum operations, there are several discriminating fashions such as  minimum-error discrimination, unambiguous discrimination, and minimax discrimination. Here we consider minimum-error discrimination between two unitary operations  ${U}_1$ and
${U}_2$.  The problem  can be reformulated into the problem
of finding a probe state $|\psi\rangle$
such that the error probability  in discriminating between the
output states $U_1|\psi\rangle$ and $U_2|\psi\rangle$ is
minimum. Denote by $P_E(U_1, U_2)$ the error probability of discriminating between $U_1$ and $U_2$. Then we have
\begin{align*}
 P_E(U_1, U_2)&=\min_{|\psi\rangle}\frac{1}{2}\left(1-\sqrt{1-4p_1p_2|\langle\psi| U_1^\dagger U_2|\psi\rangle|^2}\right)\\
 &=\frac{1}{2}\left(1-\sqrt{1-4p_1p_2\min_{|\psi\rangle}|\langle\psi| U_1^\dagger U_2|\psi\rangle|^2}\right)\\
 &=\frac{1}{2}\left(1-\sqrt{1-4p_1p_2F(U_1, U_2)^2}\right),
\end{align*}
where $p_1$ and $p_2$ are the a priori probabilities for $U_1$ and $U_2$, respectively. Here the minimum value is taken over all states $|\psi\rangle$ with $|||\psi\rangle||=1$, and  $$F(U_1, U_2)\equiv\min_{|\psi\rangle}|\langle\psi| U_1^\dagger U_2|\psi\rangle|$$ is called the fidelity of $U_1$ and $U_2$. If $F(U_1, U_2)=0$, then $ P_E(U_1, U_2)=0$, and in this case $U_1$ and $U_2$ are said to be perfectly distinguishable. Otherwise, they can  be distinguished with some error probability.

Now suppose that $U_1$ and $U_2 $ are acting on a two-qubit system $AB$, and consider how to discriminate them by LOCC. In this case, the probe state $|\psi\rangle$ should be a product state, that is, $|\psi\rangle=|\psi_A\rangle\otimes |\psi_B\rangle$. Note that for discriminating two multipartite states, it has been shown that LOCC can achieve the same distinguishability that the global operations would have. In other words, if two multipartite states are perfectly distinguishable by global operations, then they are also perfectly distinguishable by LOCC \cite{Wal2000}; if they are distinguishable with some error probability, then they can be distinguished by LOCC with the same error probability \cite{Vir2001}. Therefore, if we can find a product state  $|\psi\rangle$ such that $F(U_1, U_2)=|\langle\psi| U_1^\dagger U_2|\psi\rangle|$, then it can be asserted that   LOCC as powerful as the global operations in discriminating $U_1$ and $U_2$.

In the following, we consider discrimination between two entangling unitary operations in Eq. (\ref{Decomposition2}) (equivalently, in Eq. (\ref{Decomposition1})). Given two unitary operations $U_1$ and $U_2$ in Eq. (\ref{Decomposition2}),  the product $U_1^\dagger U_2$ also has the diagonal form of Eq. (\ref{Decomposition2}):
 \begin{equation}
U_1^\dagger U_2=\sum_{j=1}^{4}e^{-i\omega_j}|\Phi_j\rangle \langle\Phi_j|,
\label{Decomposition3}
\end{equation}
Note that it does not necessarily hold that $\omega_4\geq \omega_1\geq \omega_2\geq \omega_3$. Let $|\psi\rangle=\sum_k u_k|\Phi_k\rangle$ with $\sum_{k=1}^{4}|u_k|^2 = 1$. Then we get
 \begin{align*}
 F(U_1, U_2)&=\min_{|\psi\rangle}|\langle|\psi| U_1^\dagger U_2|\psi\rangle|\\
 &=\min\left\{ \left| \sum_{k=1}^{4}|u_k|^2
e^{-i\omega_k} \right| : \sum_{k=1}^{4}|u_k|^2 = 1 \right\}.
\label{eq:sigmax}
\end{align*}

We will show below that  the value of $F(U_1, U_2)$  can be achieved by a  product state $|\psi\rangle$ (that is, $|\psi\rangle$ satisfies $\sum_k u^2_k=0$). First, by letting $S=\left\{e^{-i\omega_k} \right\}_{k=1}^4$,  we get
 \begin{equation}conv\left(S\right)=\left\{  \sum_{k=1}^{4}|u_k|^2 e^{-i\omega_k}: \sum_{k=1}^{4}|u_k|^2 = 1 \right\},\label{conv}\end{equation}
where $conv(S)$ denotes the convex hull of  $S$. Then, $F(U_1, U_2)$ corresponds to the minimum distance from the original point $O$  to the convex hull $conv(S)$, that is,
\begin{equation*}
   F(U_1, U_2)=\min_{P\in conv\left(S\right)}||O-P||,
\end{equation*}
where it is readily seen that $O\in conv(S)$, iff $F(U_1, U_2)=0$, which means  a perfection discrimination is achievable.

\begin{figure}
\center \includegraphics[width=9cm]{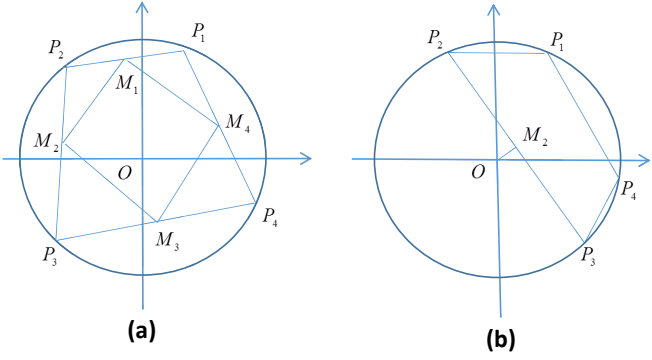}
\caption{ The convex hull of Eq. (\ref{conv}). $P_k$ corresponds to $e^{-i\omega_k}$ with $i=1,\cdots, 4$, where without loss of generality, we assume  these points on the unit circle are $P_1, P_2, P_3, P_4$ in the counter-clockwise order. $M_{i}$ denotes the midpoint of $P_i$ and $P_{(i+1)\mod 4}$  for $i=1,\cdots, 4$. In (a), the convex hull contains the original point $O$, which implies that $F(U_1, U_2)=0$, and then a perfect discrimination is achievable. In (b), the convex hull does not contain $O$. Then $F(U_1, U_2)$ is equal to the distance between $O$ and $M_2$.} \label{Fig 1}
\end{figure}

In geometry, each $e^{-i\omega_k}$ stands for a point on the unit circle in the complex plane. As shown in Fig. \ref{Fig 1}, let $P_k$ denote the point $e^{-i\omega_k}$ with $k=1,\cdots, 4$. Without loss of generality, assume the  counter-clockwise order of these points on the unit circle is $P_1, P_2, P_3, P_4$.   Denote  by $\square P_1P_2P_3P_4$ the region enclosed by the convex  polygon with endpoints $P_1, P_2, P_3, P_4$. Then  $\square P_1P_2P_3P_4$ is the convex hull $conv(S)$.

In the following we show that  the value of $F(U_1, U_2)$ is always achievable by a product state,  by discussing  two cases.

{\bf Case (i): $F(U_1, U_2)=0$.}
In this case, we have $O\in conv(S)$, or equivalently, $O\in \square P_1P_2P_3P_4$  as shown in (a) of Fig. \ref{Fig 1}.

Let $M_{i}$ denote the midpoint of $P_i$ and $P_{(i+1)\mod 4}$  with $i=1,\cdots, 4$, that is,
\begin{align*}
&M_1=\frac{1}{2}(P_1+P_2)=\frac{1}{2}(e^{-i\omega_1}+e^{-i\omega_2}), \\
& M_2=\frac{1}{2}(P_2+P_3)=\frac{1}{2}(e^{-i\omega_2}+e^{-i\omega_3}),\\
& M_3=\frac{1}{2}(P_3+P_4)=\frac{1}{2}(e^{-i\omega_3}+e^{-i\omega_4}),\\
& M_4=\frac{1}{2}(P_4+P_1)=\frac{1}{2}(e^{-i\omega_4}+e^{-i\omega_1}).
\end{align*}
First, it is not difficult to show that $O\in \square P_1P_2P_3P_4$ implies $O\in \square M_1M_2M_3M_4$, according to some geometric properties. As a result, there exist a set of positive  coefficients $\alpha_1, \alpha_2, \alpha_3, \alpha_4$ satisfying $\sum_j \alpha_j=1$ such that $\sum \alpha_j M_i=O$, that is,
\begin{align*}
&\alpha_1 \frac{(e^{-i\omega_1}+e^{-i\omega_2})}{2}+\alpha_2\frac{(e^{-i\omega_2}+e^{-i\omega_3})}{2}\\
+&\alpha_3\frac{(e^{-i\omega_3}+e^{-i\omega_4})}{2}+\alpha_4\frac{(e^{-i\omega_4}+e^{-i\omega_1})}{2}=0.
\end{align*}
It can be rewritten as
\begin{align}
&\frac{(\alpha_1+\alpha_4)}{2}e^{-i\omega_1}+\frac{(\alpha_1+\alpha_2)}{2}e^{-i\omega_2}\nonumber\\
+&\frac{(\alpha_2+\alpha_3)}{2}e^{-i\omega_3}+\frac{(\alpha_3+\alpha_4)}{2}e^{-i\omega_4}=0.
\label{eq0}\end{align}
Let
\begin{align*}
&u_1=\sqrt{(\alpha_1+\alpha_4)/2}, ~~~ u_2=i\sqrt{(\alpha_1+\alpha_2)/2},\\
&u_3=\sqrt{(\alpha_2+\alpha_3)/2}, ~~~ u_2=i\sqrt{(\alpha_3+\alpha_4)/2}.
\end{align*}
Then Eq. (\ref{eq0}) means  $\sum_k|u_k|^2e^{-i\omega_k}=0$, and one can check that $\sum_ku_k^2=0$. Therefore, we have shown that there exist a product state $|\psi\rangle=\sum_k u_k|\Phi_k\rangle$ such that $0=F(U_1, U_2)=|\langle\psi|U_1^\dagger U_2|\psi\rangle|$.

{\bf Case (ii): $F(U_1, U_2)\neq0$.}  In this case, $O\notin conv(S)$, or equivalently, $O\notin \square P_1P_2P_3P_4$  as shown in (b) of Fig. \ref{Fig 1}.  Then the minimum distance from the original
point $O$ to  $\square P_1P_2P_3P_4$  is the distance from $O$ to the line $P_2P_3$, which is equal to the distance between $O$ and the midpoint $M_2$ of $P_2$ and $P_3$.
Therefore, we obtain $$F(U_1, U_2)=|OM_2|=\|\frac{1}{2}(e^{-i\omega_2}+e^{-i\omega_3})\|.$$ Now let
\begin{align*}u_1=0, u_2=\frac{1}{\sqrt{2}}, u_3=i\frac{1}{\sqrt{2}}, u_4=0.\end{align*}  Then we have
$F(U_1, U_2)=|\langle\psi|U_1^\dagger U_2|\psi\rangle|$ for $|\psi\rangle=\sum_k u_k|\Phi_k\rangle$  with $\sum_ku_k^2=0$, that is, $F(U_1, U_2)$ is achieved by a product state.

In summary, we have shown that for any two entangling two-qubit unitary operations $U_1$ and $U_2$ in the form of Eq. (\ref{Decomposition2}), their fidelity $F(U_1, U_2)\equiv \min_{|\psi\rangle}|\langle\psi|U_1^\dagger U_2|\psi\rangle$ can be achieved by a product state $|\psi\rangle=|\psi\rangle_A \otimes|\psi\rangle_B$, and as a result,  $U_1$ and $U_2$ can be optimally discriminated by LOCC.

\section{Conclusion\label{Con}}

In this paper we have shown that  LOCC  are as powerful as  the global operations in discriminating two two-qubit entangling unitary operations without local parties, when they can be queried only once. More specifically, we have obtained:  (i) if such two unitary operations are perfectly distinguishable by global operations, then they are perfectly distinguishable by LOCC too, and (ii) if they are not perfectly distinguishable by global operations, then  LOCC can achieve the same optimal discrimination probability  as  the global operations. Therefore, LOCC suffice for   optimally distinguishing the mentioned unitary operations with only one query allowed. We hope   these discussions about this elementary case would shed some light on the following general problem: if the number of queries to unitary operations is fixed, say $k$, then when do  LOCC suffice for  optimally distinguishing bipartite unitary operations?



\bibliographystyle{plain}

\begin{thebibliography}{10}



\bibitem{Aci2001}
A. Acin, Phys. Rev. Lett.  {\bf 87}, 177901 (2001).


\bibitem{MAP2001}
G.~M. D'Ariano, P. L.~Presti, and M. G.~A. Paris,  Phys. Rev. Lett. {\bf 87}, 270404 (2001).

\bibitem{DFY2007}
R. Y. Duan, Y. Feng, and M. S. Ying, Phys. Rev. Lett. {\bf 98}, 129901 (2007).

\bibitem{Zhou2007}
X. F. Zhou, Y.S. Zhang, and G. C. Guo,
Phys. Rev. Lett.  {\bf 99}, 170401 (2007).

\bibitem{DFY2008}
R. Y. Duan, Y. Feng, and M. S. Ying, Phys. Rev. Lett. {\bf 100}, 020503 (2008).


\bibitem{LQ2008}
L. Z. Li and D. W. Qiu, Phys. Rev.  A  {\bf 77}, 032337 (2008).
\bibitem{Li2017}
L. Z. Li,  Phys. Rev.  A  {\bf 96}, 022303 (2017).

\bibitem{CAR2013} G. Chiribella,  G. M. D'Ariano,  and M. Roetteler,  New J. Phys. {\bf 15},  103019 (2013).

\bibitem{Bae2015}
J. Bae, Sci. Rep. {\bf 5}, 18270 (2015).

\bibitem{Gao2016}
T. Q. Cao, Y. H. Yang, Z. C. Zhang, G. J. Tian, F. Gao, and Q.Y. Wen,  Sci. Rep. {\bf 6}, 26696 (2016).



\bibitem{Ji2006} Z. F. Ji, Y. Feng, R. Y. Duan, and M. S. Ying,  Phys. Rev. Lett. {\bf 96}, 200401 (2006).




\bibitem{DSK2005} G. M. D'Ariano,  M. F. Sacchi,  and J. Kahn,   Phys. Rev. A {\bf 72},  052302 (2005).



\bibitem{CKTS2007}  A. Chefles,  A. Kitagawa, M. Takeoka, M. Sasaki, J. Twamley   J. Phys. A: Math. Theor. {\bf 40}, 10183 (2007).





\bibitem{Sac2005}  M. F. Sacchi,   Phys. Rev. A {\bf 71}, 062340 (2005).

\bibitem{LQ2008JPA}
L. Z. Li and D. W. Qiu,  J. Phys. A: Math. Theor. {\bf 41}, 335302 (2008).

\bibitem{WY2006}  G. M. Wang and  M. S. Ying,    Phys. Rev. A {\bf 73},  042301 (2006).


\bibitem{Wat2008} J. Watrous,  Quantum Inf. Comput. {\bf 8} (8), 819-833 (2008)

\bibitem{DFY2009} R. Y. Duan, Y. Feng, and M. S. Ying,  Phys. Rev. Lett. {\bf 103}, 210501 (2009).

\bibitem{Bry} J.-L. Brylinski and R. Brylinski, Mathematics of Quantum Computation, edited by R. Brylinski and G. Chen (CRC Press,
Boca Raton, 2002). Also see arXiv: quant-ph/0108062.

\bibitem{Horn} R. A. Horn and C. R. Johnson, {\it Topics in Matrix Analysis}
(Combridge University Press, Cambridge, 1991).

\bibitem{Kraus2000}
B.~Kraus and J.~I. Cirac,  Phys. Rev. A  {\bf 63}, 062309 (2001).

\bibitem{Wal2000} J. Walgate,  A. J. Short, L.  Hardy,  and  V. Vedral, Phys. Rev. Lett. {\bf 85}, 4972-4975 (2000).

\bibitem{Vir2001} S. Virmani,  M. F. Sacchi,  M. B. Plenio,   D. Markham,  Phys. Lett. A {\bf 288}, 62 (2001).

\end{thebibliography}

\end{document}